\documentstyle[12pt]{article}
\begin{document}
\begin{titlepage}
\begin{flushright}
 SLAC-PUB-5981\\
 January 1993\\
 T\\
 \vspace{1cm}
\end{flushright}
\begin{center}
{\Large\bf
The Heavy Quark Self-Energy in Nonrelativistic Lattice QCD}\\
\vspace{1cm}
Colin J. Morningstar\\
{\it Stanford Linear Accelerator Center}\\
{\it Stanford University, Stanford, California 94309}
\end{center}
\vspace{1cm}
The heavy quark self-energy in nonrelativistic lattice QCD
is calculated to $O(\alpha_s)$ in perturbation theory.
An action which includes all spin-independent relativistic
corrections to order $v^2$, where $v$ is the typical
heavy quark velocity, and all spin-dependent corrections
to order $v^4$ is used.  The standard Wilson action
and an improved multi-plaquette action are used for the gluons.
Results for the mass renormalization, wavefunction renormalization,
and energy shift are given; tadpole contributions are found
to be large.  A tadpole improvement scheme in which all link
variables are rescaled by a mean-field factor is also studied.
The effectiveness of this scheme in offsetting the large tadpole
contributions to the heavy quark renormalization parameters
is demonstrated.
   \vspace{5mm}
   \begin{center}
   Submitted to {\sl Physical Review D}
   \end{center}
   \vspace{5cm}
   \begin{itemize}
   \item Work supported by the Natural Sciences and Engineering
   Research Council of Canada and by the Department of Energy,
   Contract No.~DE-AC03-76SF00515.
   \end{itemize}
\end{titlepage}

\section{Introduction}

Quantum chromodynamics (QCD) has been very successful in
describing the high-energy interactions of quarks and gluons;
however, despite nearly two decades of lattice simulations,
attempts to extract from the theory reliable quantitative
information on the low-energy properties of hadrons
have not been fully successful.

Recently, a new approach has been suggested \cite{lep91}
which advocates combining effective field theory and lattice
techniques to study hadrons composed entirely of heavy quarks.
There are many good reasons for doing this.  Such hadrons are
the simplest to analyze since the quarks are nonrelativistic.
Simulations which use a nonrelativistic QCD (NRQCD) action instead
of the usual Dirac action are far more efficient since the
heavy-quark propagator can be computed as an initial-value problem
in a single sweep through the lattice.  The Dirac propagator
is a boundary-value problem which must be solved iteratively using
many sweeps.  Also, the fermion doubling problem does not occur
in the nonrelativistic theory.  Heavy quarkonia, such as the
$\psi$ and $\Upsilon$ mesons, are small so that
lattice volumes presently used are sufficiently large,
and their properties are fairly insensitive to both light-quark
vacuum polarization effects \cite{lep92}, which are very expensive
to simulate, and to heavy-quark vacuum polarization.
Simulations of heavy quarkonia do not suffer from a rapid
reduction in the signal-to-noise ratio which is a serious
shortcoming of the static approximation used to study heavy-light
systems, such as the $D$ and $B$ mesons.  Quarkonia are well
understood; this makes possible good control of systematic errors
which is necessary for precise tests of QCD.  Furthermore,
an abundance of experimental data on quarkonium states is
available to which the simulation results, such as level
splittings, decay constants, and wavefunctions, may be compared.

NRQCD is an effective cutoff field theory constructed from a set of
nonrenormalizable interactions specified solely by the
symmetries of QCD, the chosen regulator, locality, and
the accuracy desired.  It is essentially a
low-energy expansion of the Dirac theory in terms of the expectation
value $v$ of the heavy quark velocity
in a typical heavy quark hadron.
An action which includes all spin-independent relativistic
interactions suppressed by $v^2$ relative to the leading terms
and all spin-dependent corrections up to order $v^4$ has previously
been formulated using a lattice regularization scheme \cite{lep92}
and is here referred to as lattice NRQCD.  To fully define lattice
NRQCD, the coupling coefficients of the interactions appearing
in the action must be specified.  These are process and momentum
independent and are uniquely determined (for a given regulator) by
requiring that lattice NRQCD exactly reproduces the results of
continuum QCD at low energies.

Since the role of these couplings is to absorb the relativistic
effects arising from highly-ultraviolet QCD processes, one expects
that they may be computed to a good approximation using perturbation
theory, provided the quark mass $M$
is large enough.  The simplest way
to proceed is to evaluate various scattering amplitudes both in QCD
and lattice NRQCD and adjust the couplings until these amplitudes
agree to the desired order in $v$ and the QCD coupling $g$.
In this way, one obtains coupling
coefficients which are power series
in $g^2(\Lambda)$, where $\Lambda$ is the cutoff of the effective
theory.  The cutoff $\Lambda$ must be large enough so that
$O(g^2(\Lambda))$ corrections to the effective couplings
are small, making a perturbative analysis meaningful.  However,
power-law divergences generally occur, producing terms such as
$g^2(\Lambda)\Lambda/M$ which render perturbation theory useless
if $\Lambda$ is made too large.

In this paper, the lowest-order corrections to the heavy quark
self-energy in lattice NRQCD are calculated using weak-coupling
perturbation theory.  The action formulated in Ref.~\cite{lep92}
which includes spin effects is used.  Similar to an earlier
study \cite{dav92}, the mass and wavefunction renormalization
parameters required to match continuum QCD are obtained, as well
as a necessary overall energy shift.  Lattice NRQCD is briefly
described in Section~\ref{sec:nrqcd}.  The Feynman rules are
derived in Section~\ref{sec:frules} and the self-energy
calculations are presented and discussed in Section~\ref{sec:selfen}.
A tadpole improvement scheme in which all link variables are rescaled
by a mean-field factor (link variable renormalization) is also
studied.  Taking into account the mean-field corrections introduced
by this scheme, the coefficients of $g^2$ in the heavy quark
renormalization parameters are found to be small.
Section~\ref{sec:concl} offers conclusions.

\section{Lattice NRQCD}
\label{sec:nrqcd}

Standard renormalization group
techniques are used to formulate the NRQCD action.  First, since the
physics of heavy quark systems is dominated by momenta $p\sim Mv
\ll M$, where $M$ is the heavy quark mass, an ultraviolet cutoff
$\Lambda\sim M$ is introduced.  The Dirac action can then be
replaced by a nonrelativistic Schr\"odinger action in which the
quark and antiquark degrees of freedom essentially decouple.
However, new local interactions must be added in order
to compensate for the loss of the discarded
relativistic states.  Power counting rules \cite{lep92}
are used to classify these interactions according to
their estimated magnitude in a typical quarkonium state
as measured in terms of $v$.
Equipped with these rules, the interaction terms which contribute
to quarkonia physics up to some given order in $v$ may then be
enumerated.  The couplings associated with these interactions are
determined by requiring agreement between the physical results of
NRQCD and those of full QCD through the specified order in the
quark velocity.  When formulated on a lattice, NRQCD becomes a
particularly powerful tool for studying heavy quark systems.

In lattice simulations, Euclidean space correlation functions are
computed.  These may be obtained from the Minkowski theory by Wick
rotating the contour integral of the Lagrangian over time from the
real axis to the imaginary axis in a clockwise manner.  Also, the
path integration over the scalar potential must be similarly Wick
rotated at each point in space-time.  Then by defining $x^4=ix^0$,
$x_4=-ix_0$, $A^4=iA^0$, and $A_4=-iA_0$, one can again work with
real quantities.  The spatial components of all four-vectors remain
unchanged.  The path integral weight $\exp(iS)$ becomes $\exp(-S_E)$,
where $S_E$ is called the Euclidean action.  Wick rotation
transforms the Minkowski metric tensor $g_{\mu\nu}=
{\rm diag}(1,-1,-1,-1)$ into the negative of the identity matrix,
so that in Euclidean space, raising or lowering any index
introduces a sign change.

In lattice NRQCD, the quark fields are defined on the sites of a
four-dimensional hypercubic lattice with spacing $a$,
while the gauge-field degrees of freedom reside on the links
between the sites.  With each link originating at a site $x$
and terminating at a site $x+a{\hat e}_\mu$
is associated a link variable $U_\mu(x)$, which is a lattice version
of the parallel transport matrix between sites and is an element
of the Lie group associated with the gauge invariance of the theory.
A lattice gluon field $A_\mu(x)=\sum_{b=1}^8 A_\mu^b(x) T^b$
may be defined in terms of the link variables using
\begin{equation}
 U_\mu(x)=\exp\left[iagA_\mu(x+ {a \over 2}{\hat e}_\mu)\right],
\end{equation}
where $g$ is the coupling constant of the theory and
$({\hat e}_\mu)^\alpha =\delta^\alpha_\mu$.
This definition is convenient since it is simple and satisfies
$U_\mu^\dagger(x)=U_{-\mu}(x+a{\hat e}_\mu)$.
The SU(3) generators $T^b$ are Hermitian, traceless, and satisfy
${\rm Tr}(T^aT^b)={1\over 2}\delta^{ab}$ and $[T^a,T^b]=if^{abc}T^c$,
where $f^{abc}$ are the real,
fully antisymmetric structure constants.
Of course, the field $A_\mu(x)$ is not identical to the gluon field
of the continuum theory.  Let $G_\mu(x)$ represent the gluon field
defined in the continuum theory.  Under a local gauge transformation
$S(x)$, $ G_\mu(x)\longrightarrow S(x)G_\mu(x)S^\dagger(x)
-(i/g) S(x)\partial_\mu S^\dagger(x)$. The field $A_\mu(x)$ does
not transform in this way; rather,
it transforms in a very complicated
manner in order that the link variable transforms
under a local gauge transformation according to
$U_\mu(x)\longrightarrow S(x)U_\mu(x)S^\dagger(x+a{\hat e}_\mu)$.

In Minkowski space, the chromoelectric and chromomagnetic fields
are usually defined in terms of the field strength tensor
in the following manner:
$E_{(M)}^k(x) = F^{k0}(x)=F_{0k}(x)$, and
$B^k(x) = -{1\over 2}\epsilon_{klm} F^{lm}(x)$,
where $\epsilon_{123}=1$.  In Euclidean space, the magnetic field is
unchanged.  The electric field in Euclidean space is here defined by
$ E^k(x) = F^{k4}(x) = -F_{4k}(x)$,
so that $E_{(M)}^k=-iE^k$.  The Hermitian and traceless
field strength tensor is best represented \cite{man83}
by cloverleaf operators defined at the sites of the lattice:
\begin{eqnarray}
 F_{\mu\nu}(x) &=& {\cal F}_{\mu\nu}(x) -
  {1\over 3}{\rm Tr}{\cal F}_{\mu\nu}(x), \\
 {\cal F}_{\mu\nu}(x) &=& {-i\over 2a^2g}
\left( \Omega_{\mu\nu}(x)-\Omega^\dagger_{\mu\nu}(x)\right), \\
\Omega_{\mu\nu}(x) &=&  {1\over 4}\sum_{{\lbrace(\alpha,\beta)
   \rbrace}_{\mu\nu}}\!\!U_\alpha(x)U_\beta(x+a{\hat e}_\alpha)
   U_{-\alpha}(x+a{\hat e}_\alpha+a{\hat e}_\beta)U_{-\beta}
   (x+a{\hat e}_\beta),
\end{eqnarray}
with $\lbrace(\alpha,\beta)\rbrace_{\mu\nu} = \lbrace (\mu,\nu),
(\nu,-\mu), (-\mu,-\nu),(-\nu,\mu)\rbrace$ for $\mu\neq\nu$.
This representation is chosen since
it transforms as the $(1,0) \oplus (0,1)$ six-dimensional reducible
representation of the hypercubic group,
similar to the continuum case.

Covariant derivatives are replaced by forward, backward,
or symmetric differences on the lattice:
\begin{eqnarray}
a^5\Delta_\mu^{(+)}(y;x) &=&
U_\mu(y){\hat\delta}^{(4)}(x,y+a{\hat e}_\mu)
-{\hat\delta}^{(4)}(x,y),\\
a^5\Delta_\mu^{(-)}(y;x) &=&
{\hat\delta}^{(4)}(x,y)-
U^\dagger_\mu(x){\hat\delta}^{(4)}(x,y-a{\hat e}_\mu),\\
\Delta^{(\pm)} &=& {1\over 2}(\Delta^{(+)}+\Delta^{(-)}),
\end{eqnarray}
where ${\hat\delta}^{(4)}(x,y)$ denotes a four-dimensional Kronecker
$\delta$-function.  For example,
\begin{equation}
\sum_x a^5\Delta^{(+)}_\mu(y;x)\psi(x) =
U_\mu(y)\psi(y\!+\!a{\hat e}_\mu)\!-\!\psi(y).
\end{equation}
Also, the Laplacian becomes
\begin{equation}
{\bf\Delta}^{(2)}= \sum_{k=1}^3 \Delta^{(+)}_k\Delta^{(-)}_k
= \sum_{k=1}^3\Delta^{(-)}_k\Delta^{(+)}_k.
\end{equation}

The finite lattice spacing introduces systematic errors into NRQCD
which can be reduced by the addition of new interactions to the
action.  At tree level, this is most easily accomplished
by improving the components comprising the lattice action so that
they more accurately reproduce the effects of their continuum
counterparts.  An improved difference operator which reproduces
the behavior of the continuum covariant derivative through order
$a^4$ is given by
\begin{equation}
 \tilde\Delta^{(\pm)}_k = \Delta^{(\pm)}_k
- {a^2\over 6} \Delta^{(+)}_k\Delta^{(\pm)}_k\Delta^{(-)}_k.
\end{equation}
An improved lattice Laplacian is
\begin{equation}
{\bf\tilde\Delta}^{(2)}= {\bf\Delta}^{(2)}
- {a^2\over 12}\sum_{k=1}^3
\left(\Delta^{(+)}_k\Delta^{(-)}_k\right)^2.
\end{equation}
Lastly, an improved cloverleaf field strength tensor
may be defined by
\begin{eqnarray}
{\tilde F}_{\mu\nu}(x) &=&{5\over 3}F_{\mu\nu}(x)
- {1\over 6}\left[ U_\mu(x)F_{\mu\nu}(x+a{\hat
e}_\mu)U^\dagger_\mu(x)\right. \nonumber\\
&+&\left. U^\dagger_\mu(x-a{\hat e}_\mu)F_{\mu\nu}(x-a{\hat
e}_\mu)U_\mu(x-a{\hat e}_\mu)
 - (\mu \leftrightarrow\nu)\right].
\end{eqnarray}

The portion of the lattice NRQCD action containing the heavy
quark-gluon interactions may be written \cite{lep92}
\begin{eqnarray}
&&S_Q^{(n)} = a^3\sum_x\psi^\dagger(x)\psi(x) \\
&&-a^3\sum_x \psi^\dagger(x\!+\!a{\hat e}_4)\!
\left(1\!-\!{a{\tilde H}_0 \over 2n}\right)^n\!\!
\left(1\!-\!{a \delta H\over 2}\right) U^\dagger_4(x)
\left(1\!-\!{a \delta H\over 2}\right)\!
\left(1\!-\!{a{\tilde H}_0 \over 2n}\right)^n\!\!\psi(x).
\nonumber
\label{nrqcdaction}
\end{eqnarray}
This action includes all spin-independent corrections which are
suppressed by $v^2$ relative to the leading terms and all
spin-dependent interactions suppressed by factors up to
$v^4$.  The heavy quark field $\psi(x)$ is a Pauli spinor
corresponding to the two upper components of the original Dirac
field.  (An analogous action may be written for
the heavy antiquark field $\chi(x)$.)
The improved kinetic energy is given by
\begin{equation}
\tilde H_0= - {{\bf\tilde\Delta}^{(2)} \over 2M}
- {a\over 4n} {({\bf\Delta}^{(2)})^2\over 4M^2},
\end{equation}
where $n$ is a positive integer.
This operator is cleverly constructed
to produce a leading error from the lattice approximation of the
temporal derivative which can be removed by a redefinition of the
quark fields.  In this way, the heavy quark propagation
remains governed by a Schr\"odinger equation across a single time
slice without affecting energy levels, on-shell scattering
amplitudes and other physical quantities.
The parameter $n$ was introduced in Ref.~\cite{lep91}
to remove instabilities in the evolution of the quark Green's
function which occur when the temporal spacing is not small
enough to accurately treat the high-momenta modes.

The quark-gluon interactions are as follows:
\begin{equation}
\delta H = \sum_{j=1}^7 c_j\ V_j,
\end{equation}
where
\begin{eqnarray}
V_1 &=& - {({\bf\Delta}^{(2)})^2\over 8M^3}, \\
V_2 &=& {ig\over 8M^2} \bigl({\bf\Delta}^{(\pm)}\cdot {\bf E}
      - {\bf E} \cdot {\bf\Delta}^{(\pm)}  \bigl), \\
V_3 &=& - {g\over 8M^2} {\hbox{\boldmath $\sigma$}}\cdot{\bf(\tilde
        \Delta^{(\pm)}\times \tilde E - \tilde E \times
        \tilde\Delta^{(\pm)})}, \\
V_4 &=& - {g\over 2M} {\hbox{\boldmath $\sigma$}}\cdot{\bf\tilde B},
        \\
V_5 &=& - {g \over 8M^3}\lbrace{\bf\Delta}^{(2)},
        {\hbox{\boldmath $\sigma$}}
  \cdot {\bf B}\rbrace, \\
V_6 &=& - {3g\over 64M^4}\lbrace{\bf\Delta}^{(2)},
        {\hbox{\boldmath $\sigma$}}
         \cdot({\bf\Delta^{(\pm)}\!\times\! E\! -\!
         E\!\times\!\Delta^{(\pm)})}
         \rbrace, \\
V_7 &=& - {ig^2\over 8M^3} {\hbox{\boldmath $\sigma$}}
        \cdot{\bf E\times E},
\end{eqnarray}
and where the coefficients $c_j$ are
functions of $aM$ and the running coupling $\alpha_s$
in general.  At tree level, their values are all unity.
Note that the three-vector $\bf\Delta^{(\pm)}$
refers to the spatial components of the covariant four-vector
$\Delta^{(\pm)}_\mu$, while {\bf E} and {\bf B} refer to the
components $E^k$ and $B^k$, respectively.

The four-fermion contact interactions involving a quark and an
antiquark $\chi(x)$ are given by:
\begin{eqnarray}
 &&S_{\rm contact} = -g^4{a^4\over M^2} \sum_x\Bigl\lbrace
 d_1 \psi^\dagger(x)\chi(x)\chi^\dagger(x)\psi(x)
+ d_2 \psi^\dagger(x){\hbox{\boldmath $\sigma$}}
 \chi(x)\cdot\chi^\dagger(x)
 {\hbox{\boldmath $\sigma$}}\psi(x)\nonumber\\
&&+\sum_{a=1}^8\Bigl( d_3 \psi^\dagger(x)T^a\chi(x)\chi^
 \dagger(x)T^a\psi(x)
 + d_4 \psi^\dagger(x)T^a{\hbox{\boldmath $\sigma$}}
   \chi(x)\cdot\chi^\dagger(x)
 T^a{\hbox{\boldmath $\sigma$}}\psi(x)\Bigr)\Bigr\rbrace.
\end{eqnarray}
Analogous contact terms between pairs of quarks
and pairs of antiquarks also occur but only affect baryons.

The standard Euclidean gluon action is given by
\begin{equation}
 \hat S_G = a^4\sum_x {\cal L}_G(x),
\label{simpleglue}
\end{equation}
where
\begin{equation}
{\cal L}_G(x) = {1\over 2a^4g^2} \sum_{\mu\neq\nu} {\rm Tr}\bigl[
 2 - \Omega_{\mu\nu}(x)-\Omega^\dagger_{\mu\nu}(x)\bigr]
\end{equation}
is the usual single-plaquette lattice gauge field Lagrangian density.
An improved gluonic action may be written
\begin{equation}
 S_G = a^4\sum_x\bigl( {4\over 3}{\cal L}_G(x)
   -{1\over 3}{\cal L}_G^{(2\times 2)}(x)\bigr),
\label{improveglue}
\end{equation}
where ${\cal L}_G^{(2\times 2)}$ is comprised of $2\times 2$
plaquette operators and is given by:
\begin{equation}
{\cal L}^{(2\times 2)}_G(x) =
 {1\over 32a^4g^2} \sum_{\mu\neq\nu} {\rm Tr}\bigl[
 2 - \Omega^{(2)}_{\mu\nu}(x)-\Omega^{\dagger(2)}_{\mu\nu}(x)\bigr],
\end{equation}
with
\begin{eqnarray}
  &&\Omega^{(2)}_{\mu\nu}(x) =
   {1\over 4}\sum_{{\lbrace(\alpha,\beta)\rbrace}_{\mu\nu}}
   U_\alpha(x)U_\alpha(x\!+\!a{\hat e}_\alpha)
   U_\beta(x\!+\!2a{\hat e}_\alpha)
   U_\beta(x\!+\!2a{\hat e}_\alpha\! +\!a{\hat e}_\beta) \nonumber\\
  &&\times U_{-\alpha}(x\!+\!2a{\hat e}_\alpha\!+\!2a{\hat e}_\beta)
   U_{-\alpha}(x\!+\!a{\hat e}_\alpha\!
   +\!2a{\hat e}_\beta)
   U_{-\beta}(x\!+\!2a{\hat e}_\beta)U_{-\beta}
   (x\!+\!a{\hat e}_\beta),
\end{eqnarray}
and $\lbrace(\alpha,\beta)\rbrace_{\mu\nu}
 = \lbrace (\mu,\nu),(\nu,-\mu),
(-\mu,-\nu),(-\nu,\mu)\rbrace$ for $\mu\neq\nu$.
Presently, light quarks are neglected.

The lattice NRQCD action is formulated in terms of the link variables
$U_\mu(x)$ in order to preserve local gauge invariance, and as the
lattice spacing $a$ becomes small, it must tend to the action of the
continuum theory.  This can be shown at tree level in perturbation
theory using the following relationship between the link variables
and the gluon field $G_\mu(x)$ of the continuum theory:
\begin{equation}
U_\mu(x)=\exp\left[ iagA_\mu(x\!+\!\frac{a}{2}\hat e_\mu)\right]
\sim 1+iagG_\mu(x),
\end{equation}
where the lattice gluon field $A_\mu(x)=G_\mu(x)+O(a^2)$.
Beyond tree level, however, one observes that large renormalizations
are necessary to match the small $a$ limit of the lattice action
to the continuum form.  These large renormalizations stem mainly
from the higher-order powers of $agA_\mu$ which occur in the
expansion of $U_\mu$.  Such terms generate ultraviolet divergences
proportional to powers of $a^{-1}$ and so are suppressed only by
powers of $g^2$ and not $a$.

A simple gauge-invariant procedure for improving the lattice NRQCD
action by reducing the magnitudes of the renormalizations needed
to reproduce the continuum theory has been suggested:
replace all link variables $U_\mu$ that appear in the lattice
action by $U_\mu/u_0$, where $u_0$ is a parameter representing
the mean value of the link \cite{lep92}.  A gauge-invariant
definition of this mean-field parameter may be written in terms
of the mean plaquette:
\begin{equation}
u_0=\langle \frac{1}{3}{\rm Tr} U_{\rm plaq}\rangle^{1/4}.
\label{meanfieldpar}
\end{equation}
The parameter $u_0$ may be calculated using perturbation
theory or may be measured in a simulation in order to include
nonperturbative effects.  The mean-field corrections introduced
by this procedure are sometimes not small; for example,
the cloverleaf magnetic and electric fields may be nearly doubled.
Omission of such factors can result in significant underestimates
of quantities, such as spin splittings, which depend on
{\bf E} and {\bf B}.

In lattice NRQCD simulations, a value for the lattice
spacing $a$ must be chosen subject to a few constraints.
First, the cutoff $\Lambda\sim\pi/a$ must
be larger than the highest physically-relevant mass scale; thus,
$\Lambda\gg Mv$ is required.  For bottomonium, $v_b^2\sim 0.1$ and
$M_b\sim 5$ GeV so that $a\ll 2.0\ {\rm GeV}^{-1}$; for charmonium,
$v_c^2\sim 0.3$ and $M_c\sim 1.7$ GeV yielding $a\ll 3.4\ {\rm GeV}
^{-1}$.  It is also desirable to choose $\Lambda$ large enough so
that $O(g^2(\Lambda))$ corrections to the coupling coefficients
$c_j$ and $d_j$ are small.  However, NRQCD is a nonrenormalizable
effective field theory and its coupling coefficients generally
contain powers of $\Lambda/M$, causing problems if $\Lambda$ is made
too large.  Furthermore, an overly large cutoff defeats
the very purpose of NRQCD, which is to remove the scale $M$ from the
dynamics.  Hence, it is important that $\Lambda$ does not get much
larger than $M$.  Note that the requirement $aL\gg r$, where $L$ is
the extent of the lattice in number of sites and $r$ is the
root-mean-square radius of the system under study,
can be easily satisfied in present-day simulations since
$r_\psi\sim 2.4-5.0\ {\rm GeV}^{-1}$ for the $\psi$
family and $r_\Upsilon\sim 1.0-3.6\ {\rm GeV}^{-1}$ for bottomonium.

Once a suitable value for $a$ is chosen, the bare quark mass $M$ and
the bare lattice coupling $g$ must be fixed by reference to
experiment.  One very useful quantity for this purpose is the
spin-averaged $s\!-\!p$ splitting in quarkonium; due to an interplay
between the Coulombic and linear forces, this splitting is
essentially independent of $aM$.  The masses of the lowest-lying
$\psi$ and $\Upsilon$ mesons are also appropriate references.
Of course, $g$ and $M$ run with $a$ in such a way that
low-energy physical predictions do not depend on the cutoff
up to some order in $v$ and $a$. In practice,
it is much more convenient to first choose an appropriate
value for $g$, and then fix $a$ and $M$ by comparison to experiment.

\section{The Feynman Rules}
\label{sec:frules}

In a lattice gauge theory, the space of gauge transformations is
finite so that gauge-fixing is not necessary.  However,
weak-coupling perturbation theory can only be applied
if one fixes the gauge and extends the integration range of
$A_\mu^b(x)$ using the familiar Faddeev-Popov technique \cite{fad67}.
Hence, a gauge-fixing term $S_{\rm GF}$ must be added to the
NRQCD action.  The Faddeev-Popov ghost action can then be
determined from $S_{\rm GF}$ \cite{kaw81}.

To facilitate the perturbative evaluation of scattering amplitudes,
the gauge-invariant path integral measure must be expressed in terms
of the lattice gluon field \cite{bou77}:
\begin{equation}
[{\cal D}U] = \exp\left(-S_{\rm ms}\right)\left(\prod_{xb\mu}
 dA_\mu^b(x)\right),
\end{equation}
where
\begin{equation}
 S_{\rm ms} = -{1\over 2}\sum_{x\mu}{\rm Tr}\ln\left[
{2(1-\cos ag{\cal A}_\mu(x+{a\over 2}{\hat e}_\mu))\over
(ag{\cal A}_\mu(x+{a\over 2}{\hat e}_\mu))^2}\right]
\end{equation}
and ${\cal A}_\mu^{bc} = if^{bcd}A_\mu^d$.  To order $g^2$,
\begin{equation}
 S_{\rm ms} \approx {a^2g^2\over 8} \sum_{xb\mu} \left[
  A_\mu^b(x+{a\over 2}{\hat e}_\mu)\right]^2.
\end{equation}

The Feynman rules are determined by expanding the total action in
terms of the coupling $g$ and Fourier transforming into
momentum space.  The coupling coefficients are written
$c_j=1+g^2c_j^{(2)}+O(g^4)$.  Rewriting the heavy
quark-gluon action as
\begin{equation}
 S_Q^{(n)}=a^8 \sum_{xy} \psi^\dagger(y)G(y;x)\psi(x),
\end{equation}
then
\begin{equation}
S_Q^{(n)} = \int {d^4k\over (2\pi)^4}{d^4k^\prime\over
(2\pi)^4}\ {\tilde \psi}^\dagger(k^\prime){\tilde G}(k^\prime;k)
{\tilde \psi}(k),
\end{equation}
where the Fourier transforms are defined on an infinite lattice by:
\begin{eqnarray}
\psi(x) &=& \int_{\vert k_\mu\vert\leq \pi/a}
 {d^4k\over (2\pi)^4} e^{-ik\cdot x}\tilde\psi(k),\\
{\tilde G}(k^\prime;k) &=& a^8 \sum_{xy}e^{ik^\prime\cdot y
-ik\cdot x}G(y;x).
\end{eqnarray}
The perturbative expansion of ${\tilde G}(k^\prime;k)$ in terms of
the gluon fields takes the general form:
\begin{eqnarray}
&&{\tilde G}(k^\prime;k) =
(2\pi)^4\delta^{(4)}(k^\prime-k)\
a^{-1}\xi^{(0)}_{\tilde G}(k^\prime,k)\nonumber\\
&&+\sum_{\nu_1 b_1}\int d(k^\prime,k;q_1)\
{\tilde A}^{b_1}_{\nu_1}(q_1)\
\xi^{(1)}_{\tilde G}(k^\prime,k;q_1,\nu_1,b_1) \\
&&+\sum_{\nu_1\nu_2b_1b_2}\int d(k^\prime,k;q_1,q_2)\
 {\tilde A}^{b_1}_{\nu_1}(q_1){\tilde A}^{b_2}_{\nu_2}(q_2)
  a\xi^{(2)}_{\tilde G}(k^\prime,k;q_1,\nu_1,b_1;
 q_2,\nu_2,b_2) +\ \dots,\nonumber
\end{eqnarray}
where
\begin{equation}
d(k^\prime,k;q_1,\dots,q_r)
 = \left[\prod_{i=1}^r
{d^4q_i \over (2\pi)^4}\right] (2\pi)^4
\delta^{(4)}\left(k^\prime-k+\sum_{i=1}^r q_i\right)
\end{equation}
and ${\tilde A}_\mu(q)$ is the Fourier transform of
$A_\mu(x)$ defined by
\begin{equation}
A_\mu(x)=\int {d^4q\over (2\pi)^4} e^{iq\cdot x}{\tilde A}_\mu(q).
\end{equation}
Thus, the Feynman rules follow easily from the (dimensionless)
$\xi_{{\tilde G}}^{(i)}$ functions.

One of the simplest ways to compute the functions
 $\xi_{{\tilde G}}^{(i)}$
is to first calculate the Fourier transforms of the basic operators
which comprise ${\tilde G}(k^\prime;k)$
and then combine these Fourier
transforms appropriately.  The perturbative expansion of the
transform of each such operator will take the same general form as
that for ${\tilde G}(k^\prime;k)$ given above.
Of course, this general form also
applies to transform products.  Let
\begin{equation}
C_{\lbrace\alpha\beta\rbrace}(k^\prime;k) = \int {d^4p\over
 (2\pi)^4}\ A_{\lbrace\alpha\rbrace}(k^\prime;p)\
B_{\lbrace\beta\rbrace}(p;k),
\end{equation}
then the product rule for the $\xi$ functions may be expressed as
follows:
\begin{eqnarray}
 &&\xi_{C_{\lbrace\alpha\beta\rbrace}}^{(r)}
 \Bigl(k^\prime,k;\lbrace q_l,\nu_l,
 b_l\rbrace_{l=1}^r\Bigr) \\
 &&=\sum_{s=0}^r \xi_{A_{\lbrace\alpha\rbrace}}^{(s)}
\Bigl(k^\prime,k^\prime+\sum_{l=1}^s q_l;
  \lbrace q_i,\nu_i,b_i\rbrace_{i=1}^s\Bigr)
 \xi_{B_{\lbrace\beta\rbrace}}^{(r-s)}
\Bigl(k-\!\!\!\sum_{m=s+1}^r\!\! q_m,k;
 \lbrace q_j,\nu_j,b_j \rbrace_{j=s+1}^r\Bigr).
\nonumber
\end{eqnarray}
Using this rule and the additivity of the $\xi$ functions, one can
quickly build up the $\xi_{{\tilde G}}^{(i)}$
functions.  The Fourier
transforms of all the necessary basic operators are presented
in Appendix~A.

The lowest-order heavy quark propagator,
shown in Fig.~\ref{figone}(a),
is given by:
\begin{equation}
{\cal Q}^{ij}_{\alpha\beta}(k) =
a\delta^{ij}\delta_{\alpha\beta}Q(k),
\end{equation}
where $i,j$ are color indices and $\alpha,\beta$
are spin indices, and
\begin{equation}
 Q(k)= \left[
 1 - e^{ik_4a}\left(1\!+\!{\kappa_2({\bf k})^2\over M^3a^3}\right)^2
\left(1\!-\!{\kappa_2({\bf k})\over nMa}\!-\!{\kappa_4({\bf k})\over
 3nMa} \!+\!{\kappa_2({\bf k})^2\over 2n^2M^2a^2}\right)^{2n}
 \right]^{-1}
\end{equation}
with $\kappa_n({\bf k}) = \sum_{j=1}^3 \sin^n\left(k_ja/2\right)$.

In the Feynman gauge, the lowest-order gluon propagator,
shown in Fig.~\ref{figone}(b), is given by:
\begin{equation}
 {\cal D}_{\eta\nu}^{bc}(k;\lambda)
= a^2\delta^{bc}\delta_{\eta\nu}D_\nu(k;\lambda),
\end{equation}
where
\begin{equation}
 D_\nu(k;\lambda) = \biggl[{16\over 3}\biggl(\sum_{\alpha=1}^4
 \sin^2\Bigl({k_\alpha a\over 2}\Bigr)\biggr)
-{1\over 3} \cos^2\Bigl({k_\nu a\over 2}\Bigr)
\biggl(\sum_{\alpha=1}^4\sin^2\bigl(k_\alpha a\bigr)\biggr)
 +a^2\lambda^2\biggr]^{-1}.
\end{equation}
A small gluon mass $\lambda$ has been introduced to provide
an infrared cutoff.  If the simple gluon action in
Eq.~(\ref{simpleglue}) is used, the lowest-order gluon propagator
is then
\begin{equation}
 \hat{\cal D}_{\eta\nu}^{bc}(k;\lambda)
= a^2\delta^{bc}\delta_{\eta\nu}\hat D(k;\lambda),
\end{equation}
where
\begin{equation}
 \hat D(k;\lambda) = \biggl[4\biggl(\sum_{\alpha=1}^4
 \sin^2\Bigl({k_\alpha a\over 2}\Bigr)\biggr)
 +a^2\lambda^2\biggr]^{-1}.
\end{equation}

The lowest-order vertex factors corresponding to interactions
involving a heavy quark line and one to three gluons, shown
in Fig.~\ref{figone}(c), may be written:
\begin{eqnarray}
 &&{\cal V}_1(k^\prime,\alpha,i;k,\beta,j;q_1,\nu_1,b_1)
= -g\ (2\pi)^4 \delta^{(4)}\Bigl(k^\prime\!-\!k\!+\!q_1\Bigr)
 \nonumber\\
 &&\quad\quad\quad\quad
 \sum_{\mu=1}^4\sigma^\mu_{\alpha\beta}\
 T^{b_1}_{ij}\
 \zeta^{(1)}_{{\tilde G_0}\mu}\bigl(k^\prime,k;q_1,\!\nu_1\bigr), \\
&& {\cal V}_2(k^\prime,\alpha,i;k,\beta,j;q_1,\nu_1,b_1;q_2,
 \nu_2,b_2) = -ag^2\ (2\pi)^4 \delta^{(4)}\Bigl(k^\prime\!
 -\!k\!+\!q_1\!+\!q_2\Bigr) \nonumber\\
&&\quad\quad\quad\quad
 \sum_{\mu=1}^4\sigma^\mu_{\alpha\beta}\!\sum_{\tau\in
  {\cal P}_2}\!  \Bigl(T^{b_{\tau_1}}T^{b_{\tau_2}}\Bigr)_{ij}
 \zeta^{(2)}_{{\tilde G_0}\mu}\bigl(k^\prime,k;q_{\tau_1},
 \!\nu_{\tau_1}; q_{\tau_2},\!\nu_{\tau_2}\bigr), \\
&& {\cal V}_3(k^\prime,\alpha,i;k,\beta,j;q_1,\!\nu_1,\! b_1;
  q_2,\!\nu_2,\! b_2;q_3,\!\nu_3,\! b_3)
= -a^2g^3\ (2\pi)^4 \delta^{(4)}\Bigl(k^\prime\!-\!k\!+\!q_1\!
 +\!q_2\!+\!q_3\Bigr) \nonumber\\
&&\quad\quad\quad\quad
 \sum_{\mu=1}^4\sigma^\mu_{\alpha\beta}\!\sum_{\tau\in
 {\cal P}_3}\!  \Biggr[\Bigl(T^{b_{\tau_1}}T^{b_{\tau_2}}
 T^{b_{\tau_3}}\Bigr)_{ij} \zeta^{(3a)}_{{\tilde G_0}\mu}\bigl
 (k^\prime,k;q_{\tau_1},\!\nu_{\tau_1};q_{\tau_2},\!\nu_{\tau_2};
 q_{\tau_3},\!\nu_{\tau_3}\bigr)\nonumber\\
&&\quad\quad\quad\quad\quad
 +\delta_{ij}{\rm Tr}\Bigl(T^{b_{\tau_1}}T^{b_{\tau_2}}
 T^{b_{\tau_3}}\Bigr) \zeta^{(3b)}_{{\tilde G_0}\mu}
 \bigl(k^\prime,k; q_{\tau_1},\!\nu_{\tau_1};q_{\tau_2},
 \!\nu_{\tau_2}; q_{\tau_3},\!\nu_{\tau_3}\bigr)\Biggr],
\end{eqnarray}
\noindent
where ${\cal P}_r$ is the group of permutations of $r$ elements,
$\sigma^k$ are the standard Pauli spin matrices for $k=1,2,3$ and
$\sigma^4$ is a $2\times 2$ identity matrix in spin space.
The $\zeta_{\tilde G_0}$ functions are obtained from
the $\xi_{\tilde G}$ functions by neglecting the $O(g^2)$
corrections to the $c_j$ coupling coefficients.  These corrections
show up in higher-order counterterms.
Link variable renormalization in perturbation
theory leads to the addition of the following order $g^2$
counterterm, shown in Fig.~\ref{figone}(d):
\begin{eqnarray}
&& {\cal V}_{u_0}(k^\prime,\alpha,i;k,\beta,j)
= -u_0^{(2)}\ {g^2\over a} (2\pi)^4 \delta^{(4)}
\Bigl(k^\prime\!-\!k\Bigr)
  \delta_{\alpha\beta} \delta_{ij}e^{ik_4a}\nonumber\\
 &&\quad\quad\left(1\!-\!{\kappa_2({\bf k})\over nMa}\!-\!{\kappa_4
 ({\bf k})\over 3nMa}\!+\!{\kappa_2({\bf k})^2\over
2n^2M^2a^2}\right)^{2n-1}
\left(1\!+\!{\kappa_2({\bf k})^2\over M^3a^3}\right)\nonumber\\
&& \quad\quad\Biggl\lbrace
\left(1\!-\!{6\kappa_2({\bf k})\over M^3a^3}\!+\!
 {5\kappa_2({\bf k})^2\over M^3a^3}\right)
\left(1\!-\!{\kappa_2({\bf k})\over nMa}\!-\!{\kappa_4
 ({\bf k})\over 3nMa}\!+\!{\kappa_2({\bf k})^2\over
2n^2M^2a^2}\right) \nonumber\\
 &&\quad\quad +\left(1\!+\!{\kappa_2({\bf k})^2\over M^3a^3}\right)
\left({7\over 2Ma}\!-\!{4\kappa_2({\bf k})\over 3Ma}\!
-\!{4\kappa_4({\bf k})\over 3Ma}\!-\!{3\kappa_2({\bf k})
\over nM^2a^2}\!+\!{2\kappa_2({\bf k})^2\over
nM^2a^2}\right)\Biggr\rbrace,
\end{eqnarray}
\noindent
writing $u_0=1+u_0^{(2)}g^2+O(g^4)$.

The goal here is to determine the numerical values of the coupling
coefficients $c_j$ and $d_j$ and various renormalization factors
for given values of the input parameters needed in lattice
simulations, namely, the bare lattice coupling $g$ and the bare
heavy quark mass $aM$.  Since these couplings and renormalization
factors essentially absorb the relativistic effects arising from
highly-ultraviolet processes, one expects that they may be
calculated to a good approximation using weak-coupling perturbation
theory.  Using the above Feynman rules, the development of
perturbative expansions for these quantities in terms of $g$ is
straightforward.  However, there is no compelling reason to use
the bare lattice coupling for the expansion parameter.  In fact,
recent work \cite{lep92b} suggests that $g$ is a very poor choice
of expansion parameter and that much better perturbation series
result if one re-expresses the series in terms of a renormalized
coupling $g_r$ defined in terms of some physical quantity and
which runs with the relevant length scale.  This is standard practice
in continuum perturbation theory.  Of course, if calculations
could be carried out to all orders, then the choice of expansion
parameter would be immaterial.

To define a renormalized expansion parameter, a definition of the
running coupling $g_r(\mu)$ and a procedure for determining the
relevant mass scale $\mu$ must be given.  A renormalization scheme
\cite{lep92b} which defines the coupling such that the heavy quark
potential has no $g_r^4$ or higher order corrections is particularly
attractive.  This scheme is physically motivated and produces
$O(g_r^2)$ perturbative results in good agreement with simulation
results for several different quantities.  By absorbing the
higher-order contributions to the heavy quark potential into $g_r^2$,
it is hoped that higher-order contributions in other quantities
will be small.  The renormalized coupling $g_r(\mu)$ is then given
by the usual two-loop formula with $\Lambda=46.08\Lambda_{lat}$.
The scale $\mu$ is determined by averaging $\ln q^2$ over the
one-loop process of interest, where $q$ is the loop momentum.
Since the heavy quark parameters calculated here are ultraviolet
divergent quantities, one expects $\mu\approx \pi/a$.  Here,
$\bar g_r$ shall be used to denote the value $g_r(\mu)$ of the
renormalized coupling at the appropriate scale.  For example,
at $\beta=5.7$, $g_r^2(\pi/a)\approx 1.9$ and at $\beta=6.0$,
$g_r^2(\pi/a)\approx 1.7$.  Alternatively, $\bar g_r$ could simply
be added to the list of parameters which must be fixed in any
simulation by reference to experiment.

\section{The Heavy Quark Self-Energy}
\label{sec:selfen}

The heavy quark self-energy $\Sigma(p)$ may be defined by writing the
inverse quark propagator ${\cal G}^{-1}(p)$ in the form
\begin{equation}
a {\cal G}^{-1}(p)_{\alpha\beta}^{ij} =
Q^{-1}(p)\delta^{ij}\delta_{\alpha\beta}
 - a \Sigma^{ij}_{\alpha\beta}(p),
\end{equation}
where $i,j$ are color indices and $\alpha,\beta$ are spin indices.
At order $g^2$ and $v^4$ and neglecting link variable
renormalization for the moment, this self-energy is given by:
\begin{equation}
\Sigma^{ij}_{\alpha\beta}(p)=\lim_{\lambda\rightarrow 0}
g^2\delta^{ij} \delta_{\alpha\beta}
\bigl(\Sigma^{(A)}(p;\lambda)+\Sigma^{(B)}(p;\lambda)\bigr),
\end{equation}
where
\begin{equation}
a\Sigma^{(A)}(p;\lambda) = {4\over 3}a^4
\sum_{\mu,\nu=1}^4 \int {d^4k\over (2\pi)^4} Q(p-k)
 D_\nu(k;\lambda)\ \varepsilon_\mu
\bigl[\zeta_{{\tilde G_0}\mu}^{(1)}(p-k,p;k,\nu)\bigr]^2,
\label{selfdiag}
\end{equation}
corresponding to the diagram in Fig.~\ref{figtwo}(a), and
\begin{equation}
a\Sigma^{(B)}(p;\lambda) = -{4\over 3}a^4
\sum_{\nu=1}^4 \int {d^4k\over (2\pi)^4}
D_\nu(k;\lambda)
 \zeta_{{\tilde G_0} 4}^{(2)}(p,p;k,\nu;-k,\nu),
\label{tadpole}
\end{equation}
corresponding to the tadpole diagram in Fig.~\ref{figtwo}(b).
Note that $\varepsilon_\mu=(-1,-1,-1,1)$.
The following properties of the $\zeta$ functions are used
to obtain the above results:
\begin{eqnarray}
\zeta_{{\tilde G_0}\mu}^{(1)}(k^\prime,k;k-k^\prime,\nu)
 &&=\varepsilon_\mu\
\zeta_{{\tilde G_0}\mu}^{(1)}(k,k^\prime;k^\prime-k,\nu),\\
\zeta_{{\tilde G_0} j}^{(2)}(p,p;k,\nu;-k,\nu) &&=0,
 \quad\quad (j=1,2,3).
\end{eqnarray}
The self-energy is invariant under spatial reflections
$p_j\rightarrow-p_j$, for $j=1,2,3$, and transforms into its
complex conjugate under $p_4\rightarrow -p_4^\ast$.

In order to investigate $\Sigma^{(A)}(p;\lambda)$
and $\Sigma^{(B)}(p;\lambda)$ in the
neighborhood of $p=0$, the integrals in Eqs.~(\ref{selfdiag}) and
(\ref{tadpole}) must first be evaluated.
The usual initial step in the determination of such
integrals is to use the change of variables $z=\exp(\pm ik_4a)$ to
transform the integral over $k_4$ into a contour integral along the
$\vert z\vert=1$ unit circle.  Unfortunately, the complicated pole
structure of the vertex factors near $z=0$ makes difficult the
evaluation of this contour integral by the residue theory.  Due to
this fact, the simplest procedure, approximating the
four-dimensional integral by an appropriate summation as
described below, is preferred.

Because the integrand in Eq.~(\ref{tadpole}) is
a periodic analytic function of the real variables $k_\mu$
with period $2\pi$ when $\lambda>0$,
$\Sigma^{(B)}(p;\lambda)$ is numerically well approximated
by the discrete sum
\begin{equation}
a\Sigma^{(B)}(p;\lambda) \approx
-{4\over 3}{1\over N^4}\sum_k\sum_{\nu=1}^4
D_\nu(k;\lambda)
 \zeta_{{\tilde G_0} 4}^{(2)}(p,p;k,\nu;-k,\nu).
\end{equation}
In this sum, $ak_\mu=2\pi n_\mu/N$ where the
$n_\mu$ take all integer values satisfying $-N/2<n_\mu\le N/2$
for integral $N$.  The error resulting from this
approximation diminishes exponentially fast as $N\rightarrow\infty$.
However, the rate of decay of this error is directly proportional
to the mass gap $a\lambda$, creating difficulties when $a\lambda$
is small.  Fortunately, the decay rate of this error can be
dramatically increased by making the following change of variables
\cite{lus86}:
$k_\mu \rightarrow k_\mu - \alpha\sin(k_\mu)$ with
$0\le\alpha < 1$.  This transformation maintains periodicity and
effectively increases the mass gap so that the approximation
\begin{equation}
a\Sigma^{(B)}(p;\lambda) \approx
-{4\over 3}{1\over N^4}\sum_k
\sum_{\nu=1}^4 \varrho(k)
D_\nu\bigl(s(k);\lambda\bigl)
\zeta_{{\tilde G_0} 4}^{(2)}
 \bigl(p,p;s(k),\nu;-s(k),\nu\bigr),
\label{tadpolesum}
\end{equation}
where $s_\mu(k)=k_\mu - \alpha\sin(k_\mu)$ and
$\varrho(k) = \prod_{\mu=1}^4[1-\alpha\cos(k_\mu)]$,
converges much more quickly as $N$ is increased.
The parameter $\alpha$ should be chosen so as to maximize the
effective mass gap: $\alpha = {\rm sech}(u)$, where $u$ satisfies
$a\lambda \approx u - {\rm tanh}(u)$.

The above procedure is sufficient for evaluating
$\Sigma^{(B)}(p;\lambda)$ as long
as the gluon mass $a\lambda$ is not set too small. However,
the pole in the quark propagator is problematical when evaluating
$\Sigma^{(A)}(p;\lambda)$ near $p=0$.  To circumvent this, the
contour for the $ak_4$ integral, which runs along the real axis
from $-\pi$ to $\pi$ except near the pole,
can be continuously deformed into a contour consisting of three
line segments passing through the points $-\pi\rightarrow
-\pi-ia\lambda/2\rightarrow\pi-ia\lambda/2\rightarrow\pi$.
This contour is chosen for the following two reasons:  for $p\sim 0$,
the distance of closest approach to any pole is a maximum;
and the contributions from the segments of the contour running
parallel to the imaginary axis cancel due to the periodicity of the
integrand.  $\Sigma^{(A)}(p\sim 0;\lambda)$ can then be accurately
obtained using the following approximation:
\begin{eqnarray}
&&a\Sigma^{(A)}(p\!\sim\! 0;\lambda) \approx {4\over 3}{1\over N^4}
\sum_k\sum_{\mu,\nu=1}^4  \varrho(k)
Q\bigl(p-r(k)\bigr)\nonumber\\
&&\quad\times D_\nu\bigl(r(k);\lambda\bigl) \varepsilon_\mu
\Bigl[\zeta_{{\tilde G_0}\mu}^{(1)}\bigl(p-r(k),p;r(k),
 \nu\bigr)\Bigr]^2,
\label{selfdiagsum}
\end{eqnarray}
where $r_\mu(k)=k_\mu - \alpha\sin(k_\mu)-ia\lambda\delta_{4,\mu}/2$
and $ak_\mu=2\pi n_\mu/N$ with the
$n_\mu$ taking all integer values satisfying $-N/2<n_\mu\le N/2$.
Also, $\alpha = {\rm sech}(y)$ and $a\lambda/2 \approx y -
{\rm tanh}(y)$. In practice, the approximations in
Eqs.~(\ref{tadpolesum}) and (\ref{selfdiagsum}) are applied using
increasing values of $N$ until sufficient convergence is observed;
typically, $N=20$ is adequate.  Note that for values of $p$
satisfying $p_x=p_y=p_z$, the number of terms which must be
independently evaluated in these sums may be dramatically reduced
by exploiting the invariance of the summands under interchange of
any two spatial components of $k$.

The evaluation of $\Sigma^{(A)}(p\sim 0;\lambda)$ and
$\Sigma^{(B)}(p\sim 0;\lambda)$ using Eqs.~(\ref{selfdiagsum}) and
(\ref{tadpolesum}) is feasible only if the gluon mass
$\lambda$ is not set too small.  However, the limits of these
functions as $\lambda\rightarrow 0$ are actually required,
necessitating the use of an extrapolation procedure.
The expected behavior of these quantities as $\lambda$ tends to zero
may be expressed as an asymptotic expansion of the form
\begin{equation}
\Sigma(p;\lambda)
\mathop{\sim}\limits_{\lambda\rightarrow 0+}
\sum_{m=0}^\infty \left(b^{(0)}_m+b^{(1)}_m\ln
a^2\lambda^2\right)(a\lambda)^{m},
\end{equation}
where the coefficient $b^{(1)}_0$ is known from the continuum theory
since the infrared divergence is insensitive to the ultraviolet
regulator.  If one neglects the $b^{(1)}_m$ terms for $m>0$, then
polynomial extrapolation using Neville's algorithm may be applied
to $f_0(p;\lambda)=\Sigma(p;\lambda)-b_0^{(1)}\ln a^2\lambda^2$.
The extrapolation can be improved by applying Neville's algorithm
to the function $f_1(p;\lambda)=f_0(p;\lambda)-a\lambda
\partial_{(a\lambda)}\Sigma(p;\lambda)+2b_0^{(1)}$ which has no
$a\lambda\ln a^2\lambda^2$ term.  Polynomial extrapolation of
$f_2(p;\lambda)=f_1(p;\lambda)+(a^2\lambda^2/2)
\partial_{(a\lambda)}^2\Sigma(p;\lambda)+b_0^{(1)},$ which does
not suffer from $a^2\lambda^2\ln a^2\lambda^2$ and $a\lambda\ln
a^2\lambda^2$ effects, is another method.  Since the gluon
mass appears only in the gluon propagator,
the derivatives of $\Sigma(p;\lambda)$ with respect to $a\lambda$
can be exactly and efficiently taken.  In practice, polynomial
extrapolation of all three functions $f_0(p;\lambda)$,
$f_1(p;\lambda)$, and $f_2(p;\lambda)$ is done using between six
and twelve values of the gluon mass lying in the range
$0.15\le a\lambda \le 0.8$, and agreement among the results is
verified. In calculating $\Sigma^{(B)}(p;\lambda\rightarrow 0)$,
one finds that only even powers
of $a\lambda$ occur in its asymptotic
expansion.  In this case, the accuracy of the extrapolation can be
increased by explicitly excluding
the odd powers in the extrapolating
polynomial.  Neville's algorithm is then applied to the
functions $f_0(p;\lambda)$,
$\tilde f_1(p;\lambda)=f_0(p;\lambda)-a^2\lambda^2
\partial_{(a^2\lambda^2)}\Sigma(p;\lambda)+b_0^{(1)}$ and
$\tilde f_2(p;\lambda)=\tilde f_1(p;\lambda)+(a^4\lambda^4/2)
\partial_{(a^2\lambda^2)}^2\Sigma(p;\lambda)+b_0^{(1)}/2$.
Uncertainties in the extrapolated values are estimated
by examining the spread of values in the Neville table and by
comparing the results obtained using different sets of gluon
mass values and using the different functions described above.

The small $v$ expansion of the zeroth-order heavy quark inverse
propagator, keeping only those terms which are suppressed by no
more than $v^2$ relative to the leading terms, is given by:
\begin{equation}
Q^{-1}(p) \approx -ip_4a + {{\bf p}^2a^2\over 2Ma}+{p_4^2a^2\over 2}
+ {ip_4{\bf p}^2a^3\over 2Ma}
- {({\bf p}^2)^2a^4\over 8M^3a^3}
- {({\bf p}^2)^2a^4\over 8M^2a^2} + \dots,
\end{equation}
recalling that $p_4$ is of order $v^2$.  The on-mass-shell
quark then satisfies the following dispersion relation:
\begin{equation}
 \omega_0({\bf p}) \approx i\Bigl({{\bf p}^2\over 2M}
- {({\bf p}^2)^2\over 8M^3} + \dots \Bigr),
\end{equation}
where $\omega_0({\bf p})$ denotes the value of the fourth
component of the contravariant quark four-momentum $p^\mu$
at the pole of the zeroth-order propagator.  This dispersion
relation agrees exactly with the continuum form from full QCD
to this order in $v$.

In the ${\overline {MS}}$ renormalization scheme, the inverse
propagator to order $g^2$ for full QCD in continuum Minkowski
space has the form
\begin{equation}
 (1-C_{cont}g^2) \left( {\not \!p} - M(1+\delta_{cont}g^2)\right)
\end{equation}
where
\begin{eqnarray}
C_{cont} &=& {1\over 12\pi^2}\left( -4\!+\!\ln(M^2/\mu^2)\!+\!
 2\ln(M^2/\lambda^2)\right), \\
\delta_{cont} &=& {1\over 12\pi^2} \left( 4+ 3\ln(\mu^2/M^2)\right).
\end{eqnarray}
Note that $\mu$ is the mass scale introduced by dimensional
regularization and $\lambda$ is the gluon mass regulating
the infrared divergence.  Thus, the sole effect of the order $g^2$
corrections is to renormalize the quark field and mass.  If lattice
NRQCD is to reproduce the low-energy physical predictions of
full QCD, then the order $g^2$ corrections to the heavy quark
propagator in lattice NRQCD must
also do no more than renormalize the heavy quark field and mass to
the appropriate order in $v$.

Explicit calculation shows that the
heavy quark self-energy in lattice
NRQCD has a small $v$ representation of the following form:
\begin{equation}
 a\Sigma(p) \approx g^2\Bigl\lbrace \Omega_0 -i p_4a \Omega_1
+ {{\bf p}^2a^2\over 2Ma}\Omega_2 + \dots\Bigr\rbrace,
\end{equation}
retaining only radiative corrections to the lowest-order terms
in $v$ as specified by the power counting rules of Ref.~\cite{lep92}.
The on-mass-shell quark now satisfies a dispersion relation given by
\begin{equation}
 \omega_0({\bf p})\approx i\Bigl( - g^2\Omega_0 /a
+ {{\bf p}^2\over 2M} \left(1+g^2\Omega_1-g^2\Omega_2\right)
- {({\bf p}^2)^2\over 8M^3} + \dots \Bigr).
\end{equation}
Defining $M_r=Z_m M$, where $Z_m=1-g^2\Omega_1+g^2\Omega_2$ is the
mass renormalization factor, and $\bar p_4=p_4-ig^2\Omega_0/a$,
the inverse propagator for small $v$ may be written:
\begin{equation}
 a{\cal G}^{-1}(p)\approx Z_\psi \left(
 -i \bar p_4a + {{\bf p}^2a^2\over 2M_ra} + {\bar p_4^2a^2\over 2}
 + {i \bar p_4{\bf p}^2a^3\over 2M_ra}
- {({\bf p}^2)^2a^4\over 8M_r^3a^3}
- {({\bf p}^2)^2a^4\over 8M_r^2a^2} + \dots\right),
\end{equation}
where $Z_\psi=1-g^2(\Omega_0+\Omega_1)$ is the wavefunction
renormalization parameter.  Thus, the addition of a counterterm
which shifts the energy by an overall amount $\bar g_r^2\Omega_0/a$
is needed in order to match the low-energy physical predictions
of lattice NRQCD with those of QCD.  Alternatively, one could
simply shift the energies obtained in simulations using the
action in Eq.~\ref{nrqcdaction} by an amount $\bar g_r^2\Omega_0/a$
for each heavy quark.

A more convenient set of renormalization parameters may be obtained
by defining $Z_\psi=1-g^2C$, $Z_m=1+g^2B$,
and $\bar p_4=p_4+ig^2A/a$.
The parameters $A$, $B$, and $C$ can then be calculated using
$A=-\Omega_0$, $B=\Omega_2-\Omega_1$, and $C=\Omega_0+\Omega_1$,
where $\Omega_0=a\Sigma(0)/g^2$, $\Omega_1=ia\partial_{p_4a}
\Sigma(0)/g^2$, and $\Omega_2=2Ma^2\partial_{{\bf p}^2a^2}\Sigma(0)
/g^2$.  Due to the complexity of the NRQCD vertex factors, the
derivatives in these expressions must be taken numerically.
Four- and five-point formulas are applied in the differentiation;
only points which satisfy $p_x=p_y=p_z$ are used since the
self-energy can be computed much more quickly at such points.

Results for the energy shift parameter $A$ and the mass
renormalization parameter $B$ are presented in Tables~\ref{tableA}
and \ref{tableB}, respectively.  Both $A$ and $B$ are gauge
invariant and infrared finite.  The wavefunction renormalization
parameter $C$ has an infrared divergence of the form
$-(\ln a^2\lambda^2)/6\pi^2$ as the gluon mass is
taken to zero.  This divergence is cancelled in physical
quantities by an infrared divergence occurring in the
quark-gluon vertex correction.   Values for the infrared-finite
portion of $C$ are given in Table~\ref{tableC}.
In these tables, the contributions from the quark-gluon loop
and tadpole diagrams are given separately.  Results using
the simple and the improved gluonic actions are presented.
Values for the stability parameter $n$ are chosen
such that the pole in the quark propagator $Q(p)$ falls on the
same side of the real axis in the complex $p_4$ plane for all
allowed values of ${\bf p}$ and tends to move farther away from this
real axis as $\vert {\bf p}\vert$ increases to its allowed maximum.
For $aM\ge 2$, $n$ is set to unity; for $1\le aM< 2$, $n=2$ is
used.  These values of $n$ ensure the stability of the evolution
equation for the quark Green's function.

Contributions from the quark-gluon loop graph to the energy
shift parameter $A$ are small and decrease in magnitude as $aM$
decreases.  At large $aM$, there is little difference between
the shifts obtained using the simple and improved gluonic
actions.  The tadpole contributions are large and
contain power-law divergences which grow in magnitude as $aM$
is decreased.  Since high-momentum modes are more strongly damped
in the improved gluon propagator, the ultraviolet-divergent
tadpole terms are appreciably smaller in the case of the
improved gluon action.  These total downward shifts in the energy
are nearly the same as those obtained in Ref.~\cite{dav92} which
used a much simpler heavy quark action. Calculations of $A$ using
$c_j=0$ for $j=3,4,5,6,7$ reveal that contributions to this
parameter from the spin-dependent interactions are small.

Contributions to the mass renormalization parameter $B$ from the
quark-gluon loop diagram are very small and do not vary
proportionately with $aM$.  For all values of $aM$, there is
little difference between the results obtained using
the simple and the improved gluon actions.  The tadpole
contributions are again large, growing in magnitude as $aM$
is decreased.  The tadpole terms in the case of the improved
action are slightly smaller and contributions to this parameter
from the spin-dependent interactions are small.
The total values for $B$ obtained here are appreciably larger
than the order $g^2$ corrections to the mass renormalization
calculated in Ref.~\cite{dav92} (see Ref.~\cite{davnote}).

The tadpole diagrams do not contribute to the heavy quark
wavefunction renormalization parameter $C$.  The values
for the infrared-finite portion of $C$ are very small and
become increasingly negative as $aM$ is decreased.  There
is little difference between the results obtained using the
simple and the improved gluonic actions.  The magnitudes
of the wavefunction renormalization corrections are much
smaller than those obtained in Ref.~\cite{dav92}
(see Ref.~\cite{davnote}).

The contributions to the energy shift and mass renormalization
parameters from the link variable renormalization counterterm
are given by:
\begin{eqnarray}
\delta A &=& u_0^{(2)}\left( 1 + {7\over 2Ma}\right), \\
\delta B &=& u_0^{(2)}\left({2\over 3}-{1\over 4nMa}
 +{3\over M^2a^2}\right).
\end{eqnarray}
No change in the wavefunction renormalization occurs.
As previously stated, the purpose of link variable renormalization
is to enhance the similarities between lattice and continuum
gauge-field operators, especially those depending on the
cloverleaf electric and magnetic fields.  Consequently, the
counterterms introduced by this renormalization offset
the large tadpole contributions which commonly afflict lattice
perturbation theory, offering a means of improving its convergence.
As shown in Table~\ref{tableD}, the order $g^2$ corrections to
the heavy quark renormalization parameters are small once the
mean-field corrections are taken into account.  In this table,
the value $u_0^{(2)}=-0.083$ obtained by evaluating
Eq.~\ref{meanfieldpar} in perturbation theory for the simple
gluonic action is used.  For $\bar g_r^2\sim 2$ and
$a\sim 1 {\rm GeV}^{-1}$, the $b$ quark receives approximately a
$1\%$ lowest-order correction to its mass and the $c$ quark
receives about a $7\%$ mass correction.

\section{Conclusion}
\label{sec:concl}

Lattice NRQCD is an effective field theory which promises to
make possible high-precision numerical studies of heavy quark
systems.  It is essentially a low-energy expansion of QCD
in terms of the mean velocity of the heavy quarks
in a typical heavy-quark hadron.  To fully define lattice
NRQCD, the coupling strengths of its interactions must be specified.
These are determined by requiring that lattice NRQCD reproduces
the low-energy physical results of continuum QCD.
Since the role of these couplings is to absorb the relativistic
effects arising from highly-ultraviolet QCD processes, one expects
that they may be computed to a good approximation using perturbation
theory, provided the quark mass $M$ is large enough.

The heavy quark self-energy in nonrelativistic lattice QCD
was calculated to $O(\alpha_s)$ in perturbation theory.
An action which includes all spin-independent relativistic
corrections to order $v^2$ and all spin-dependent corrections
to order $v^4$ was used.  The standard Wilson action
and an improved multi-plaquette action were used for the gluons.
Results for the mass and wavefunction renormalization
and an overall energy shift were obtained.  Contributions
from the quark-gluon loop graph were found to be very small;
however, the tadpole contributions were large.
The values of these parameters will
be needed in future numerical simulations of quarkonium.
The effective couplings will also be needed;
calculation of these quantities is in progress.

A tadpole improvement scheme in which all link variables are
rescaled by a mean-field factor $u_0$ was also applied in
perturbation theory.  The main purpose of this link variable
renormalization was to enhance the similarities between lattice
and continuum gauge-field operators, especially those depending
on the cloverleaf electric and magnetic fields.  An important
consequence of this scheme was a significant offsetting of the
large tadpole contributions to the heavy quark renormalization
parameters.  Using a perturbative approximation to the mean
plaquette for $u_0$, the tadpole-improved heavy quark
renormalization parameters were shown to be small.
This scheme offers a means of improving the
convergence properties of lattice perturbation theory.

\vspace{1cm}
\centerline{\Large\bf Acknowledgments}

\noindent
I would like to thank G.~Peter~Lepage for many useful
conversations.
This work was supported by the Natural Sciences and Engineering
Research Council of Canada and by the Department of
Energy, Contract No.~DE-AC03-76SF00515.
\vspace{1cm}
\begin{flushleft}{\Large\bf Appendix}\end{flushleft}

The momentum-space representations of various components of the
lattice NRQCD action are presented in this Appendix.
Below, $\zeta$ functions are introduced and are defined by
\begin{equation}
 \xi_A^{(r)}(k^\prime,k;q_1,\!\nu_1,\! b_1;\dots;q_r,\!\nu_r,
\! b_r)=g^r \zeta_A^{(r)}(k^\prime,k;q_1\!,\!\nu_1;\dots;
q_r\!,\!\nu_r)\ T^{b_1}\cdots T^{b_r}.
\end{equation}
Also note that $\bar\delta_{\mu\nu}$ is an ``anti-Kronecker delta
function'' and is trivial when $\mu=\nu$ and unity otherwise.

In momentum space, the link variable becomes
\begin{eqnarray}
&&U_\mu(k^\prime;k) =
a^4\sum_x e^{i(k^\prime-k)\cdot x}e^{-ik_\mu a}
  \exp\left[iagA_\mu(x+{a\over 2}{\hat e}_\mu)\right] \nonumber\\
&&= (2\pi)^4\delta^{(4)}(k^\prime-k)
 e^{-ia(k^\prime_\mu+k_\mu)/2} +iag\int d(k^\prime,k;q)
 e^{-ia(k^\prime_\mu+k_\mu)/2}{\tilde A}_\mu(q)\nonumber\\
&&-{a^2g^2\over 2} \int d(k^\prime,k;q_1,q_2)
 e^{-ia(k^\prime_\mu+k_\mu)/2}{\tilde A}_\mu(q_1)
 {\tilde A}_\mu(q_2) \nonumber\\
&&-{ia^3g^3\over 6} \int d(k^\prime,k;q_1,q_2,q_3)
 e^{-ia(k^\prime_\mu+k_\mu)/2}{\tilde A}_\mu(q_1)
 {\tilde A}_\mu(q_2){\tilde A}_\mu(q_3)+ \cdots.
\end{eqnarray}

The momentum-space representation of $\Delta_\mu^{(\pm)}$ may be
written:
\begin{eqnarray}
\zeta^{(0)}_{a\Delta^{(\pm)}_\mu}(k^\prime,k)&=&
  -i\sin\bigl [{a\over 2}(k^\prime+k)_\mu\bigr ], \nonumber\\
\zeta^{(1)}_{a\Delta^{(\pm)}_\mu}(k^\prime,k;q_1,\!\nu_1)&=&
  ia\cos\bigl [{a\over 2}(k^\prime+k)_\mu\bigr ]\
\delta_{\mu,\nu_1}, \nonumber\\
\zeta^{(2)}_{a\Delta^{(\pm)}_\mu}(k^\prime,k;q_1,\!\nu_1;
 q_2,\!\nu_2)&=&
  i{a^2\over 2}\sin\bigl [{a\over 2}(k^\prime+k)_\mu
  \bigr ]\ \delta_{\mu,\nu_1} \delta_{\mu,\nu_2}, \\
\zeta^{(3)}_{a\Delta^{(\pm)}_\mu}(k^\prime,k;q_1,\!\nu_1;q_2,
 \!\nu_2;q_3,\!\nu_3)&=&
  -i{a^3\over 6}\cos\bigl [{a\over 2}(k^\prime+k)_\mu\bigr
  ]\ \delta_{\mu,\nu_1} \delta_{\mu,\nu_2}
 \delta_{\mu,\nu_3}.\nonumber
\end{eqnarray}
The improved symmetric derivative has the following momentum-space
representation:
\begin{eqnarray}
&&\zeta^{(0)}_{a\tilde\Delta^{(\pm)}_\mu}(k^\prime,k)=
  i\left(-{4\over 3}\sin\bigl [{a\over 2}(k^\prime+k)_\mu\bigr ]
  +{1\over 6}\sin\bigl [a(k^\prime+k)_\mu\bigr ]\right), \\
&&\zeta^{(1)}_{a\tilde\Delta^{(\pm)}_\mu}
  (k^\prime,\!k;\!q_1,\!\nu_1)=
  i{a\over 3}\delta_{\mu,\nu_1}\left(4\cos\bigl [{a\over 2}
  (k^\prime+k)_\mu\bigr ] \!-\!\cos\bigl [a(k^\prime\!+\!k)_\mu
  \bigr ]\cos({a\over 2} q_{1\mu}) \right), \nonumber\\
&&\zeta^{(2)}_{a\tilde\Delta^{(\pm)}_\mu}(k^\prime,k;q_1,\!
  \nu_1;q_2,\!\nu_2)= ia^2\delta_{\mu,\nu_1} \delta_{\mu,\nu_2}
  \left({2\over 3}\sin\bigl [{a\over 2}(k^\prime+k)_\mu\bigr ]\
 \right.\nonumber\\
&&\quad\quad \left.-{1\over 3}\sin\bigl [a(k^\prime+k)_\mu\bigr ]
  \cos({a\over 2} q_{1\mu}) \cos({a\over 2} q_{2\mu})
  \!-\!{1\over 6}\cos\bigl [a(k^\prime+k)_\mu\bigr ]
  \sin\bigl [{a\over 2}(q_1-q_2)_\mu\bigr ]  \right), \nonumber\\
&&\zeta^{(3)}_{a\tilde\Delta^{(\pm)}_\mu}(k^\prime,\!k;q_1,\!
  \nu_1;q_2,\!\nu_2;q_3,\!\nu_3)=
  ia^3\delta_{\mu,\nu_1} \delta_{\mu,\nu_2}\delta_{\mu,\nu_3}
  \left(-{2\over 9}\cos\bigl [{a\over 2}(k^\prime+k)_\mu\bigr ]
  \ \right.\nonumber\\
&&\quad\quad \!+\!\left.{1\over 18}\!
  \cos\bigl [a(k^\prime\!+\!k)_\mu\bigr ]
  \cos\bigl [{a\over 2} (k^\prime\!-\!k)_\mu\bigr ]
  \!+\!{1\over 6}\! \cos\bigl [a(k^\prime\!
  +\!k\!+\!{1\over 2}q_1\!-\!
  {1\over 2}q_3)_\mu\bigr ] \cos({a\over 2} q_{2\mu}) \right).
\nonumber
\end{eqnarray}
The Fourier transform of the lattice Laplacian may be written:
\begin{eqnarray}
\zeta^{(0)}_{a^2\Delta^{(2)}}(k^\prime,k)&=&
  -4\sum_{j=1}^3 \sin^2\bigl [{a\over 4}(k^\prime+k)_j\bigr ],
 \nonumber\\
\zeta^{(1)}_{a^2\Delta^{(2)}}(k^\prime,k;q_1,\!\nu_1)&=&
  2a\sin\bigl [{a\over 2}(k^\prime+k)_{\nu_1}\bigr ]\
 \bar\delta_{4,\nu_1}, \cr
\zeta^{(2)}_{a^2\Delta^{(2)}}(k^\prime,k;q_1,
  \!\nu_1;q_2,\!\nu_2)&=&
  -a^2\cos\bigl [{a\over 2}(k^\prime+k)_{\nu_1}
   \bigr ]\ \delta_{\nu_1,\nu_2}
   \bar\delta_{4,\nu_1}, \\
\zeta^{(3)}_{a^2\Delta^{(2)}}(k^\prime,k;q_1,\!
  \nu_1;q_2,\!\nu_2;q_3,\!\nu_3)&=&
  -{a^3\over 3}\sin\bigl [{a\over 2}(k^\prime+k)_{\nu_1}
   \bigr ]\ \delta_{\nu_1,\nu_2}
   \delta_{\nu_2,\nu_3}\bar\delta_{4,\nu_1}.\nonumber
\end{eqnarray}
Similarly for the improved Laplacian:
\begin{eqnarray}
&&\zeta^{(0)}_{a^2\tilde\Delta^{(2)}}(k^\prime,k)=
-4\sum_{j=1}^3 \left(\sin^2\bigl [{a\over 4}(k^\prime+k)_j\bigr ]+
{1\over 3} \sin^4\bigl [{a\over 4}(k^\prime+k)_j\bigr ]\right), \\
&&\zeta^{(1)}_{a^2\tilde\Delta^{(2)}}(k^\prime,\!k;\!q_1,\!\nu_1)=
  {a\over 3}\bar\delta_{4,\nu_1}\left(8\sin\bigl
  [{a\over 2}(k^\prime+k)_{\nu_1}\bigr ]
  \!-\!\sin\bigl [a(k^\prime\!+\!k)_{\nu_1}\bigr
   ]\cos({a\over 2} q_{1\nu_1}) \right), \nonumber\\
&&\zeta^{(2)}_{a^2\tilde\Delta^{(2)}}(k^\prime,k;
   q_1,\!\nu_1;q_2,\!\nu_2)=
  a^2\delta_{\nu_1,\nu_2}\bar\delta_{4,\nu_1}
 \left(-{4\over 3}\cos\bigl [{a\over 2}(k^\prime+k)_{\nu_1}
  \bigr ]\ \right.\nonumber\\
&&\quad\quad \left.+{1\over 3}\cos\bigl
   [a(k^\prime\!+\!k)_{\nu_1}\bigr ]
  \cos({a\over 2} q_{1\nu_1})\cos({a\over 2} q_{2\nu_1})
  \!-\!{1\over 6}\sin\bigl [a(k^\prime\!+\!k)_{\nu_1}\bigr ]
  \sin\bigl [{a\over 2}(q_1\!-\!q_2)_{\nu_1}\bigr ] \right),
  \nonumber\\
&&\zeta^{(3)}_{a^2\tilde\Delta^{(2)}}(k^\prime,k;
  q_1,\!\nu_1;q_2,\!\nu_2;q_3,\!\nu_3)=
  a^3\delta_{\nu_1,\nu_2} \delta_{\nu_2,\nu_3}\bar\delta_{4,\nu_1}
  \left(-{4\over 9}\sin\bigl [{a\over 2}
  (k^\prime+k)_{\nu_1}\bigr ]\ \right.\nonumber\\
&&\quad\quad\left.+{1\over 6}\sin\bigl
   [a(k^\prime\!+\!k\!+\!{1\over 2}q_1\!
  -\!{1\over 2}q_3)_{\nu_1}\bigr ]\cos({a\over 2} q_{2\nu_2})\!+\!
  {1\over 18}\sin\bigl [a(k^\prime\!+\!k)_{\nu_1}\bigr ]
  \cos\bigl [{a\over 2}(k^\prime\!-\!k)_{\nu_1}\bigr ]
   \right).\nonumber
\end{eqnarray}

Another important operator is the cloverleaf field strength tensor.
Its Fourier transform is given by
\begin{eqnarray}
&&F_{\mu\nu}(k^\prime;k)
= a^4 \sum_x e^{i(k^\prime-k)\cdot x} F_{\mu\nu}(x) \nonumber\\
&&= {i\over a}\sum_b \int  d(k^\prime,k;q)\
\bigl [f_{\mu\nu}^A(q)\ {\tilde A}^b_\nu(q)
- f_{\nu\mu}^A(q)\ {\tilde A}^b_\mu(q) \bigr ]\ T^b\nonumber\\
 &&+ ig \sum_{b_1b_2}\int d(k^\prime,k;q_1,q_2)\ T^{b_1} T^{b_2}
 \Biggl\lbrace f_{\mu\nu}^B(q_1,q_2)\
  {\tilde A}^{b_1}_\nu(q_1){\tilde A}^{b_2}_\nu(q_2)\nonumber\\
 &&-\!f_{\nu\mu}^B(q_1\!,\!q_2) {\tilde A}^{b_1}_\mu(q_1)
  {\tilde A}^{b_2}_\mu(q_2)
 \!+\!f_{\mu\nu}^C(q_1\!,\!q_2) {\tilde A}^{b_1}_\mu(q_1)
  {\tilde A}^{b_2}_\nu(q_2)
 \!-\!f_{\nu\mu}^C(q_1\!,\!q_2) {\tilde A}^{b_1}_\nu(q_1)
  {\tilde A}^{b_2}_\mu(q_2)  \Biggr\rbrace\nonumber\\
 &&+ iag^2\sum_{b_1b_2b_3}\int d(k^\prime,k;q_1,q_2,q_3)\
   \Bigl(T^{b_1}T^{b_2}T^{b_3}-{1\over 3}{\rm Tr}
   T^{b_1}T^{b_2}T^{b_3}\Bigr)\nonumber\\
   &&\Biggl\lbrace  -f_{\mu\nu}^F(q_1,q_2,q_3)\
   {\tilde A}^{b_1}_\nu(q_1){\tilde A}^{b_2}_\mu(q_2)
   {\tilde A}^{b_3}_\nu(q_3) +f_{\mu\nu}^D(q_1,q_2,q_3)\
   {\tilde A}^{b_1}_\mu(q_1){\tilde A}^{b_2}_\mu(q_2)
   {\tilde A}^{b_3}_\mu(q_3) \nonumber\\
   &&-f_{\nu\mu}^D(q_1,q_2,q_3)\
   {\tilde A}^{b_1}_\nu(q_1){\tilde A}^{b_2}_\nu(q_2)
   {\tilde A}^{b_3}_\nu(q_3) -f_{\nu\mu}^E(q_1,q_2,q_3)\
   {\tilde A}^{b_1}_\mu(q_1){\tilde A}^{b_2}_\mu(q_2)
   {\tilde A}^{b_3}_\nu(q_3) \nonumber\\
   &&-f_{\nu\mu}^E(q_3,q_2,q_1)\
   {\tilde A}^{b_1}_\nu(q_1){\tilde A}^{b_2}_\mu(q_2)
   {\tilde A}^{b_3}_\mu(q_3) +f_{\mu\nu}^E(q_1,q_2,q_3)\
   {\tilde A}^{b_1}_\nu(q_1){\tilde A}^{b_2}_\nu(q_2)
   {\tilde A}^{b_3}_\mu(q_3) \nonumber\\
   &&+f_{\mu\nu}^E(q_3,q_2,q_1)\
   {\tilde A}^{b_1}_\mu(q_1){\tilde A}^{b_2}_\nu(q_2)
   {\tilde A}^{b_3}_\nu(q_3) +f_{\nu\mu}^F(q_1,q_2,q_3)\
   {\tilde A}^{b_1}_\mu(q_1){\tilde A}^{b_2}_\nu(q_2)
   {\tilde A}^{b_3}_\mu(q_3) \Biggr\rbrace \nonumber\\
   &&+ {\cal O}(g^3)
\end{eqnarray}
where
\begin{eqnarray}
 f_{\mu\nu}^A(q) &=& \sin(aq_\mu)\cos({a\over 2} q_\nu),\nonumber\\
f_{\mu\nu}^B(q_1,q_2) &=& \cos\bigl [{a\over 2}
 (q_1+q_2)_\mu\bigr ]\sin\bigl [{a\over 2}(q_1+q_2)_\nu\bigr ]
   \sin\bigl [{a\over 2}(q_1-q_2)_\mu\bigr ], \nonumber\\
f_{\mu\nu}^C(q_1,q_2) &=& {1\over 2}\bigl [
 \cos({a\over 2} q_{1\mu})\cos(aq_{1\nu}+{a\over 2} q_{2\nu})
+ \cos({a\over 2} q_{2\nu})\cos(aq_{2\mu}+
  {a\over 2} q_{1\mu}) \nonumber\\
&+& \cos(aq_{2\mu}+{a\over 2} q_{1\mu})
  \cos(aq_{1\nu}+{a\over 2} q_{2\nu})
- \cos({a\over 2} q_{1\mu})\cos({a\over 2}
  q_{2\nu})\bigr ], \nonumber\\
f_{\mu\nu}^D(q_1,q_2,q_3) &=& \cos\bigl
  [{a\over 2}(q_1+q_2+q_3)_\mu\bigr ]
 \Biggl\lbrace {1\over 6} \sin\bigl
  [a(q_1+q_2+q_3)_\nu\bigr ]\nonumber\\
 &-&\cos\bigl [{a\over 2}(q_1+q_2+q_3)_\nu\bigr ]
 \cos\bigl [{a\over 2} (q_1-q_3)_\nu\bigr ]\sin({a\over 2}
  q_{2\nu})\Biggr\rbrace, \nonumber\\
f_{\mu\nu}^E(q_1,q_2,q_3) &=& \cos\bigl
 [{a\over 2}(q_1\!+\!q_2\!+\!q_3)_\mu\bigr ]
 \!\cos\bigl [{a\over 2}(\!q_1\!+\!q_2\!+\!q_3\!)_\nu\bigr ]
\!\cos\bigl [{a\over 2}(\!q_1\!+\!q_2\!)_\mu
 \bigr ]\sin({a\over 2} q_{3\nu}\!) \nonumber\\
 &+& {1\over 2}\cos\bigl [a(q_2+{1\over 2}q_3)_\mu\bigr ]
  \sin\bigl [{a\over 2}(q_1+q_2)_\nu\bigr ], \\
f_{\mu\nu}^F(q_1,q_2,q_3) &=& \cos\bigl
 [{a\over 2}(q_1+q_2+q_3)_\mu\bigr ]
   \sin\bigl [a({1\over 2}q_1+q_2+{1\over 2}q_3)_\nu\bigr ]
   \cos\bigl [{a\over 2}(q_1-q_3)_\mu\bigr ].\nonumber
\end{eqnarray}
Using ${\tilde A}^b_\mu(q)^\ast={\tilde A}^b_\mu(-q)$,
one can easily check that
$F^b_{\mu\nu}(k^\prime;k)^\ast=F^b_{\mu\nu}(-k^\prime;-k)$.
Furthermore, $F_{\mu\nu}(k^\prime;k)=-F_{\nu\mu}(k^\prime;k)$, as
required.  Note the absence of a ${\rm Tr}T^{b_1}T^{b_2}$ term
in the order $g$ coefficient.

The Fourier transform of the improved cloverleaf field strength
tensor has the same form as that for $F_{\mu\nu}$,
but the functions $f_{\mu\nu}^A,\dots$ must be
replaced by the following functions:
\begin{eqnarray}
 &&\tilde f_{\mu\nu}^A(q) = {1\over 3}
   \bigl(5-\cos(aq_\mu)-\cos(aq_\nu)\bigr)
   f^A_{\mu\nu}(q), \nonumber\\
&&\tilde f_{\mu\nu}^B(q_1,q_2) = {1\over 3}\Bigl\lbrace
  \bigl(5-\cos\bigl [a(q_1+q_2)_\mu\bigr ]-\cos\bigl
   [a(q_1+q_2)_\nu\bigr ]\bigr)
   f^B_{\mu\nu}(q_1,q_2) \nonumber\\
   &&\quad\quad-\sin\bigl
   [a(q_1+{1\over 2}q_2)_\nu\bigr ]f^A_{\mu\nu}(q_1)
    +\sin\bigl [a({1\over 2}q_1+q_2)_\nu\bigr
   ]f^A_{\mu\nu}(q_2)\Bigr\rbrace,\nonumber\\
&&\tilde f_{\mu\nu}^C(q_1,q_2) = {1\over 3}\Bigl\lbrace
  \bigl(5-\cos\bigl [a(q_1+q_2)_\mu\bigr ]-
   \cos\bigl [a(q_1+q_2)_\nu\bigr ]\bigr)
   f^C_{\mu\nu}(q_1,q_2) \nonumber\\
   &&\quad\quad+\sin\bigl
   [a(q_1+{1\over 2}q_2)_\nu\bigr ]f^A_{\nu\mu}(q_1)
    +\sin\bigl [a({1\over 2}q_1+q_2)_\mu\bigr ]
   f^A_{\mu\nu}(q_2)\Bigr\rbrace,\nonumber\\
&&\tilde f_{\mu\nu}^D(q_1,\!q_2,\!q_3) = {1\over 3}\Bigl\lbrace
  \bigl(5\!-\!\cos\bigl [a(q_1\!+\!q_2\!+\!q_3)_\mu\bigr ]\!-\!\cos
   \bigl [a(q_1\!+\!q_2\!+\!q_3)_\nu\bigr ]\bigr)
   f^D_{\mu\nu}(q_1,q_2,q_3) \nonumber\\
   &&\quad\quad+\sin\bigl [a(q_1+q_2+{1\over 2}q_3)_\mu\bigr
   ]f^B_{\nu\mu}(q_1,q_2)
    -\sin\bigl [a({1\over 2}q_1+q_2+q_3)_\mu\bigr
   ]f^B_{\nu\mu}(q_2,q_3)\nonumber\\
  &&\quad\quad-{1\over 2}\cos\bigl [a(q_1+{1\over 2}q_2
  +{1\over 2}q_3)_\mu\bigr ]f^A_{\nu\mu}(q_1)
  -{1\over 2}\cos\bigl [a({1\over 2}q_1+{1\over 2}q_2+q_3)_\mu
  \bigr ]f^A_{\nu\mu}(q_3)\nonumber\\
  &&\quad\quad+\cos\bigl
   [a({1\over 2}q_1+q_2+{1\over 2}q_3)_\mu\bigr
   ]f^A_{\nu\mu}(q_2)
   \Bigr\rbrace ,\nonumber\\
&&\tilde f_{\mu\nu}^E(q_1,\!q_2,\!q_3) = {1\over 3}\Bigl\lbrace
  \bigl(5\!-\!\cos\bigl [a(q_1\!+\!q_2\!+\!q_3)_\mu\bigr ]\!-\!\cos
   \bigl [a(q_1\!+\!q_2\!+\!q_3)_\nu\bigr ]\bigr)
   f^E_{\mu\nu}(q_1,q_2,q_3) \nonumber\\
   &&\quad\quad-\sin\bigl [a(q_1+q_2+{1\over 2}q_3)_\mu\bigr
  ]f^B_{\mu\nu}(q_1,q_2)
    -\sin\bigl [a({1\over 2}q_1+q_2+q_3)_\nu\bigr
  ]f^C_{\nu\mu}(q_2,q_3)\nonumber\\
   &&\quad\quad-{1\over 2}\cos\bigl [a({1\over 2}q_1
   +{1\over 2}q_2+q_3)_\nu\bigr ]
    f^A_{\nu\mu}(q_3)\Bigr\rbrace ,\nonumber\\
&&\tilde f_{\mu\nu}^F(q_1,\!q_2,\!q_3) = {1\over 3}\Bigl\lbrace
   \bigl(5\!-\!\cos\bigl [a(q_1\!+\!q_2\!+\!q_3)_\mu\bigr ]\!-\!\cos
   \bigl [a(q_1\!+\!q_2\!+\!q_3)_\nu\bigr ]\bigr)
   f^F_{\mu\nu}(q_1,q_2,q_3) \nonumber\\
  &&\quad\quad-\sin\bigl [a(q_1+q_2+{1\over 2}q_3)_\nu\bigr
   ]f^C_{\nu\mu}(q_1,q_2)
   -\sin\bigl [a({1\over 2}q_1+q_2+q_3)_\nu\bigr
   ]f^C_{\mu\nu}(q_2,q_3)\nonumber\\
  &&\quad\quad-\cos\bigl [a({1\over 2}q_1+q_2
    +{1\over 2}q_3)_\nu\bigr ]
    f^A_{\nu\mu}(q_2)\Bigr\rbrace.
\end{eqnarray}

\newpage
\centerline{\Large\bf Figures}
\begin{figure}[h]
\caption{Various Feynman diagram elements.  A curly line represents
a gluon; a double solid line indicates a heavy quark.  (a) Heavy
quark propagator; (b) gluon propagator; (c) lowest-order vertices
involving a heavy quark line and $r$ gluons; (d) the $O(g^2)$
counterterm from link variable renormalization.}
\label{figone}
\end{figure}
\begin{figure}[h]
\caption{Two Feynman diagrams which contribute to the heavy quark
self-energy.  A curly line denotes a gluon; a double solid line
denotes a heavy quark.}
\label{figtwo}
\end{figure}
\vfill
\newpage
\begin{table}[h]
\caption[a]{The energy shift parameter $A$ for various values
of the product of the bare heavy quark mass $M$ and the
lattice spacing $a$.  The contribution to $A$ from the
quark-gluon loop diagram of Fig.~\ref{figtwo}(a) is denoted by
$A_i(A)$ for the improved gluon action of Eq.~(\ref{improveglue})
and by $A_s(A)$ for the simple
gluon action of Eq.~(\ref{simpleglue}).
The contribution from the tadpole diagram of Fig.~\ref{figtwo}(b)
is denoted by $A_i(B)$ and $A_s(B)$ for the improved and
simple gluon actions, respectively.  For $aM\ge 2$, the stability
parameter $n$ is set to unity; for $1\le aM<2$, $n=2$ is used.
Extrapolation uncertainties are no larger than $\pm 0.0001$.}
\begin{center}
\begin{tabular}{r@{\hspace{15mm}}r@{\hspace{15mm}}r
@{\hspace{15mm}}r@{\hspace{15mm}}r}
\hline\hline
$aM$ & $A_i(A)$ & $A_i(B)$ & $A_s(A)$ & $A_s(B)$ \\
\hline
 5.00 & 0.0417 & 0.1361 & 0.0414 & 0.1688 \\
 4.75 & 0.0407 & 0.1387 & 0.0403 & 0.1719 \\
 4.50 & 0.0397 & 0.1415 & 0.0391 & 0.1754 \\
 4.25 & 0.0385 & 0.1446 & 0.0377 & 0.1793 \\
 4.00 & 0.0372 & 0.1480 & 0.0363 & 0.1835 \\
 3.75 & 0.0358 & 0.1519 & 0.0347 & 0.1883 \\
 3.50 & 0.0342 & 0.1562 & 0.0329 & 0.1937 \\
 3.25 & 0.0325 & 0.1611 & 0.0309 & 0.1998 \\
 3.00 & 0.0306 & 0.1667 & 0.0288 & 0.2067 \\
 2.75 & 0.0284 & 0.1732 & 0.0263 & 0.2147 \\
 2.50 & 0.0261 & 0.1807 & 0.0237 & 0.2241 \\
 2.25 & 0.0234 & 0.1896 & 0.0208 & 0.2351 \\
 2.00 & 0.0206 & 0.2002 & 0.0176 & 0.2482 \\[2mm]
 1.90 & 0.0199 & 0.2046 & 0.0168 & 0.2536 \\
 1.80 & 0.0185 & 0.2101 & 0.0154 & 0.2603 \\
 1.70 & 0.0172 & 0.2160 & 0.0139 & 0.2677 \\
 1.60 & 0.0158 & 0.2226 & 0.0124 & 0.2759 \\
 1.50 & 0.0143 & 0.2300 & 0.0108 & 0.2850 \\
 1.40 & 0.0129 & 0.2383 & 0.0092 & 0.2952 \\
 1.30 & 0.0114 & 0.2476 & 0.0076 & 0.3068 \\
 1.20 & 0.0100 & 0.2584 & 0.0061 & 0.3201 \\
 1.10 & 0.0086 & 0.2708 & 0.0047 & 0.3354 \\
 1.00 & 0.0075 & 0.2850 & 0.0036 & 0.3530 \\
\hline\hline
\end{tabular}
\end{center}
\label{tableA}
\end{table}
\newpage
\begin{table}[h]
\caption[b]{The heavy quark mass renormalization parameter $B$
for various values of the product of the bare heavy quark mass
$M$ and the lattice spacing $a$.  The contribution to $B$ from the
quark-gluon loop diagram of Fig.~\ref{figtwo}(a) is denoted by
$B_i(A)$ for the improved gluon action of Eq.~(\ref{improveglue})
and by $B_s(A)$ for the simple
gluon action of Eq.~(\ref{simpleglue}).
The contribution from the tadpole diagram of Fig.~\ref{figtwo}(b)
is denoted by $B_i(B)$ and $B_s(B)$ for the improved and
simple gluon actions, respectively.  For $aM\ge 2$, the stability
parameter $n$ is set to unity; for $1\le aM<2$, $n=2$ is used.
Extrapolation uncertainties are no larger than $\pm 0.0001$
for the tadpole values and $\pm 0.0002$ for the contributions
from the quark-gluon loop diagram.}
\begin{center}
\begin{tabular}{r@{\hspace{15mm}}r@{\hspace{15mm}}r
@{\hspace{15mm}}r@{\hspace{15mm}}r}
\hline\hline
$aM$ & $B_i(A)$ & $B_i(B)$ & $B_s(A)$ & $B_s(B)$ \\
\hline
 5.00 & $-0$.0024 & 0.0556 & $-0$.0030 & 0.0697 \\
 4.75 & $-0$.0016 & 0.0568 & $-0$.0021 & 0.0712 \\
 4.50 & $-0$.0008 & 0.0582 & $-0$.0011 & 0.0728 \\
 4.25 &  0.0000 & 0.0599 & $-0$.0002 & 0.0748 \\
 4.00 &  0.0009 & 0.0619 & 0.0008 & 0.0771 \\
 3.75 &  0.0018 & 0.0642 & 0.0019 & 0.0798 \\
 3.50 &  0.0028 & 0.0670 & 0.0030 & 0.0832 \\
 3.25 &  0.0039 & 0.0704 & 0.0042 & 0.0872 \\
 3.00 &  0.0049 & 0.0747 & 0.0054 & 0.0923 \\
 2.75 &  0.0059 & 0.0801 & 0.0065 & 0.0987 \\
 2.50 &  0.0069 & 0.0871 & 0.0076 & 0.1070 \\
 2.25 &  0.0077 & 0.0963 & 0.0085 & 0.1181 \\
 2.00 &  0.0080 & 0.1091 & 0.0089 & 0.1334 \\[2mm]
 1.90 &  0.0076 & 0.1175 & 0.0083 & 0.1438 \\
 1.80 &  0.0077 & 0.1248 & 0.0084 & 0.1525 \\
 1.70 &  0.0077 & 0.1332 & 0.0084 & 0.1626 \\
 1.60 &  0.0076 & 0.1431 & 0.0083 & 0.1746 \\
 1.50 &  0.0073 & 0.1550 & 0.0079 & 0.1889 \\
 1.40 &  0.0068 & 0.1692 & 0.0073 & 0.2061 \\
 1.30 &  0.0060 & 0.1868 & 0.0064 & 0.2274 \\
 1.20 &  0.0048 & 0.2089 & 0.0051 & 0.2541 \\
 1.10 &  0.0032 & 0.2373 & 0.0032 & 0.2885 \\
 1.00 &  0.0010 & 0.2752 & 0.0007 & 0.3345 \\
\hline\hline
\end{tabular}
\end{center}
\label{tableB}
\end{table}
\newpage
\begin{table}[h]
\caption[c]{The infrared-finite portion of the
heavy quark wavefunction renormalization parameter $C$
for various values of the product of the bare heavy quark mass $M$
and the lattice spacing $a$.  The infrared-finite
contribution to $C$ from the
quark-gluon loop diagram of Fig.~\ref{figtwo}(a) is denoted by
$C_i$ for the improved gluon action of Eq.~(\ref{improveglue})
and by $C_s$ for the simple gluon action of Eq.~(\ref{simpleglue}).
The tadpole diagram of Fig.~\ref{figtwo}(b) does not contribute
to this parameter.  For $aM\ge 2$, the stability
parameter $n$ is set to unity; for $1\le aM<2$, $n=2$ is used.
Extrapolation uncertainties are no larger than $\pm 0.0001$.}
\begin{center}
\begin{tabular}{r@{\hspace{15mm}}r@{\hspace{15mm}}r}
\hline\hline
$aM$ & $C_i$\hspace{4mm} & $C_s$\hspace{4mm} \\
\hline
  5.00 &  0.0032 &  0.0029 \\
  4.75 &  0.0018 &  0.0014 \\
  4.50 &  0.0002 & $-0$.0004 \\
  4.25 & $-0$.0015 & $-0$.0023 \\
  4.00 & $-0$.0033 & $-0$.0044 \\
  3.75 & $-0$.0054 & $-0$.0067 \\
  3.50 & $-0$.0077 & $-0$.0092 \\
  3.25 & $-0$.0102 & $-0$.0120 \\
  3.00 & $-0$.0131 & $-0$.0152 \\
  2.75 & $-0$.0163 & $-0$.0187 \\
  2.50 & $-0$.0199 & $-0$.0226 \\
  2.25 & $-0$.0239 & $-0$.0270 \\
  2.00 & $-0$.0285 & $-0$.0319 \\[2mm]
  1.90 & $-0$.0300 & $-0$.0335 \\
  1.80 & $-0$.0322 & $-0$.0358 \\
  1.70 & $-0$.0345 & $-0$.0382 \\
  1.60 & $-0$.0369 & $-0$.0408 \\
  1.50 & $-0$.0395 & $-0$.0435 \\
  1.40 & $-0$.0422 & $-0$.0464 \\
  1.30 & $-0$.0450 & $-0$.0493 \\
  1.20 & $-0$.0480 & $-0$.0524 \\
  1.10 & $-0$.0511 & $-0$.0555 \\
  1.00 & $-0$.0542 & $-0$.0587 \\
\hline\hline
\end{tabular}
\end{center}
\label{tableC}
\end{table}
\newpage
\begin{table}[h]
\caption[d]{Tadpole improvement of the energy shift parameter $A$
and mass renormalization parameter $B$ for various values of the
product of the bare mass $M$ and the lattice spacing $a$.
Results are given for the simple gluon action only.
The renormalization parameters without tadpole improvement
are denoted by $A_s$ and $B_s$; the improved parameters
are denoted by $\tilde A_s$ and
$\tilde B_s$.  The perturbative value $u_0^{(2)}=-0.083$ is
used for the mean-field parameter.}
\begin{center}
\begin{tabular}{r@{\hspace{15mm}}r@{\hspace{15mm}}r
@{\hspace{15mm}}r@{\hspace{15mm}}r}
\hline\hline
$aM$ & $A_s$\hspace{2mm} & $\tilde A_s$\hspace{2mm}
& $B_s$\hspace{2mm} & $\tilde B_s$\hspace{2mm} \\
\hline
 5.00 & 0.210 & $ 0$.069 & 0.067 & 0.006 \\
 4.00 & 0.220 & $ 0$.064 & 0.078 & 0.012 \\
 3.00 & 0.236 & $ 0$.056 & 0.098 & 0.022 \\
 2.00 & 0.266 & $ 0$.038 & 0.142 & 0.035 \\
 1.70 & 0.282 & $ 0$.028 & 0.171 & 0.036 \\
 1.30 & 0.314 & $ 0$.008 & 0.234 & 0.039 \\
 1.00 & 0.357 & $-0$.017 & 0.335 & 0.041 \\
\hline\hline
\end{tabular}
\end{center}
\label{tableD}
\end{table}

\newpage
\begin{thebibliography}{99}
\bibitem{lep91}
   B.A.~Thacker and G.~Peter~Lepage,
   Phys.\ Rev.\ D {\bf 43}, 196 (1991).
\bibitem{lep92}
   G.~Peter Lepage, Lorenzo Magnea, Charles Nakhleh, Ulrika Magnea,
   and Kent Hornbostel, Phys.~Rev.~D~{\bf 46}, 4052 (1992).
\bibitem{dav92}
   C.~T.~H.~Davies and B.~A.~Thacker,
   Phys.\ Rev.\ D {\bf 45}, 915 (1992).
\bibitem{man83}
   J.~Mandula, G.~Zweig, and J.~Govaerts,
   Nucl.\ Phys. {\bf B228}, 109 (1983).
\bibitem{fad67}
   L.D.~Faddeev and V.N.~Popov,
   Phys.\ Lett.\ {\bf 25B}, 29 (1967).
\bibitem{kaw81}
   Hikaru Kawai, Ryuichi Nakayama, and Koichi Seo,
   Nucl.\ Phys. {\bf B189}, 40 (1981).
\bibitem{bou77}
   David~G.~Boulware,
   Ann.\ Phys. {\bf 56}, 140 (1970).
\bibitem{lep92b}
   G.~Peter~Lepage and Paul~B.~Mackenzie, NSF-ITP-90-227
   (unpublished).
\bibitem{lus86}
   M.~L\"uscher and P.~Weisz,
   Nucl.\ Phys. {\bf B266}, 309 (1986).
\bibitem{davnote}
   Note that in terms of the parameters $A$, $B$, $A_S$, and $Z$
   defined in Ref.~\cite{dav92}, the order $g^2$ corrections to the
   mass and wavefunction renormalization are given by $aA+B$ and
   $Z-aA_S$, respectively.  See also Table~II in Ref.~\cite{lep92}.
\end{thebibliography}
\end{document}